# Epistemology mediating conceptual knowledge: Constraints on data accessible from interviews

Michael C. Wittmann, University of Maine, Orono ME 04469-5709, wittmann@umit.maine.edu
Rachel E. Scherr, University of Maryland, College Park MD 20742-4111, escherr@physics.umd.edu

A student's epistemological stance (be it knowledge as memorized information, knowledge from authority, or knowledge as invented stuff) may constrain that student from reasoning in productive ways while also shaping the inferences a researcher can make about how that student reasons about a particular phenomenon. We discuss both cases in the context of an individual student interview on charge flow in wires. In the first part of the interview, her focus on memorized knowledge prevents the researcher from learning about her detailed reasoning about current. In the second part of the interview, her focus on constructed knowledge provides the researcher with a picture of her reasoning about the physical mechanisms of charge flow.

## Introduction

Students enter our classes with a set of epistemological tools regarding how they know what they know and why they believe what they say.[1,2,3] The interplay between expectations and attitudes toward learning and conceptual reasoning, however, is not well understood. In this paper, we claim that a student's epistemological stance toward a specific topic in physics can mediate both the student's thinking and what a researcher can see of the student's thinking.

We describe an interview in which a student passes through two distinct phases of reasoning. In the first, her speech reflects memorized knowledge and shows little evidence of conceptual reasoning about the physics. In the second, her speech reflects ideas that she develops herself, and shows much greater evidence of conceptual reasoning. By comparing her behavior in the two phases of the interview, we as researchers gain insight into both her conceptual reasoning and her epistemological stance. We also make a case for how a student's epistemological mode in a specific context affects what a researcher can learn about a student's reasoning.[4]

## An Interview on Current and Conductivity

In a previous paper,[5,6] one author (MCW) and his collaborators described a set of interviews in which junior and senior science and engineering students were asked to compare charge flow in different materials placed in different environments. Students were given a simple apparatus (a battery with two leads) to which they could attach steel wire, copper wire, rubber bands, wood, or doped and undoped semiconductors. Students also had to consider how the current would change if the circuits were placed in different temperature environments.

We expected advanced undergraduate students to recognize that current is the flow of charged particles through a circuit. In the semi-classical atomic model, the charged particles flow through a semi-fixed atomic lattice; when heated, the lattice vibrates, so the particles have a shorter mean free path and move more slowly, thus reducing the current in a heated conducting element. We expected that some students, being electrical engineers, would be able to describe the band structure of materials and discuss conducting and valence bands.

Students commonly gave one (or both) of two descriptions during the interviews. The first was the "electron pull description," in which electrons are pulled off individual atoms before they can flow. Students often used this description to account for differences between conductors and insulators. Some said current would be greater in hot wires because the added energy would make it easier to pull electrons from atoms. The second was the "atomic jump description," in which electrons jump from atom to atom, having to be pulled off again in order to jump to the next atom. Students often used this description to reason that a battery was necessary to keep pulling electrons off the atoms and thus keep charge flowing.

## Conceptual and Epistemological Issues in a Single Interview

Sarah (an alias) was an excellent advanced physics student with high grades. We divide her interview into two distinct phases.[7] In the first phase, her answers are brief and lack detail. In the second, she develops the electron pull description to account for the effect of temperature on charge flow. Her epistemological stance toward answering questions about current directly affects the inferences that can be made when trying to make sense of the data in the interview.

### Phase 1: Memorized knowledge limits access to data

In the first part of the interview, the interviewer's insight into Sarah's reasoning is constrained by her nearly exclusive use of memorized knowledge and the recitation of ideas from other classes with no further explanation. The way she describes the source of and applicability of her knowledge directly influences her responses, so that little can be learned about her conceptual knowledge. It is unclear, and impossible to know based on the data, if she has a weak conceptual model of electrical current during this phase of the interview, or if there is nothing for the interviewer to uncover.

Two interview excerpts illustrate the manner in which her conceptual reasoning is constrained by her epistemological stance. In the first, Sarah has just discussed insulators as being so dense that electrons cannot pass through them. Then, the following dialog occurs.

Interviewer: Take something like Styrofoam. What's going on with Styrofoam? What category [conductor or insulator?] does that fall into?
Sarah: [Insulating]
I: Okay. Why?
S: Memorized it!
I: Memorized it. Okay. What property of Styrofoam might lead to that?
S: Back to the little density thing .. I don't really know.
I: Okay. It's pretty .. Styrofoam, I mean, it's not terribly dense.
S: Right. I don't really know. Something inhibits the electrons from moving quickly.

In this excerpt, Sarah's guiding epistemology toward current and charge flow is clearly stated by her comment, "Memorized it!" She answers without giving a mechanism, and apparently without thinking about the mechanism. Because her answer is memorized, she "[doesn't] really know" why electron flow is different in Styrofoam than in a metal wire, only that it is different. The interviewer cannot discern what physical mechanism, if any, she uses to distinguish between insulators, conductors, and semiconductors.

In a later segment of the interview, the interviewer asks Sarah to describe current in more detail. She has previously guessed (with no explanation) that wires at a higher temperature will carry a larger current than wires at a low temperature. In the process of explaining (ex post facto) her reasoning, she talks about free electrons.

I: So why is it that these electrons are free to move?
S: Well, maybe there are bound electrons in here [the atomic lattice describing the wire], and electrons flowing out of the battery around the loop.
I: Okay. What's 'bound electrons'?

S: Electrons that are bound to these structures [atoms in a lattice]
I: Okay. What determines if it's bound or not?
S: The energy of the structure.
I: How do you mean?
S: I have no idea! That's organic chemistry!

Sarah again shows that she depends on outside knowledge with her reference to "organic chemistry" as a body of knowledge that would explain the source of free and bound electrons. She says, "I have no idea!" as to the physical mechanism that is in play.

These two excerpts are consistent with her responses throughout most of phase 1 of the interview. Sarah presents her knowledge of electrical conduction as something memorized, something handed down from authority. Her approach directly influences the manner in which she presents her conceptual knowledge to the interviewer and constrains the inferences that can be made to observe what, if anything, she knows of electric charge flow.

## Phase 2: Constructed knowledge provides insight into reasoning

In the second phase of the interview, Sarah's answers begin to include explanations and evidence of knowledge construction, which suggests that her epistemological approach to electrical conduction has shifted to one of "knowledge as invented stuff."[2,3] By showing how she puts together elements of her thinking into new ideas, Sarah gives the interviewer insight into her reasoning. Thus, her epistemological stance acts as a driver for the information that the researcher is seeking in this interview.

The second interview excerpt above ended with Sarah focusing on "organic chemistry" as a source of information. The following dialog came immediately after:

S: I have no idea! That's organic chemistry!

I: Again! It could be. Any explanation you find.
S: Well, yeah, the electrons are bound to these molecules, and it takes certain energies to tear them away.
I: Okay, and so once they are torn away, then they're free to move?
S: Yep.
I: Okay. What tears them away? What determines that?
S: … seems like it should be an easy question .. I assume just the battery … the power supply.

When the interviewer prompts her to think of "any explanation you find," the nature of Sarah's explanations changes. From this point forward, Sarah invents ideas based on other pieces of her understanding. She brings up a new idea (the electron pull description, in essence), and builds a response to what "seems like … an easy question." Previously, she had spoken of "electrons flowing out of the battery around the loop,"[8] but now she speaks of the battery "tear[ing] them away" from the atomic lattice. Later, the interviewer asked her to revisit her answers.

I: So when the electron is moving through [the wire] and the temperature … of the one wire is much higher than the temperature of the other, [which has the higher current]?
S: So it's got a higher energy, so it would be like more electrons can be torn away from these structures, so more current will flow through the higher temperature one.
S, later: … Yeah, I guess higher temperature. I didn't know why I didn't get the connection between the higher energy and the electrons being freed, but that's what I'm going with.

Sarah's epistemological stance at this stage of the interview is that creating a description of the physics based on her ideas is valuable and useful, and "that's what [she's] going with." She commits to the description she developed, and proceeds to use it in later parts of the inter-view to explain her thinking on other topics. In the process, the interviewer gains insight into her reasoning about the physics: flowing electrons must be torn from the atomic lattice, and it is easier to tear electrons in high temperature wires from the lattice, creating more current.

## Discussion

Sarah's primary epistemological mode in each different phase of the interview directly affects the level of conceptual reasoning that is visible in her responses. In phase 1 of the interview, she focuses on memorized knowledge and the interviewer is unable to see the possible quality of her thinking. Her epistemological stance toward questions about current during this part of the interview acts as a filter. In phase 2, she presents ideas constructed from other elements of reasoning, and thereby shows her conceptual reasoning in explicit detail.

Because our knowledge of Sarah's conceptual understanding is limited by her epistemological stance, we do not know how well she understands the physics at any given time. We do not claim that she does not understand the physics in phase 1, in which she focuses on memorized, outside knowledge. Instead, we simply cannot see her reasoning in any detail. We also do not claim that the reasoning she shows in phase 2 of the interview is new and different from before, only that we are now aware of it. Thus, her epistemological stance may affect the level of her own conceptual reasoning, and definitely affects our knowledge as researchers of her thinking.

Based on our results, we question the level of our understanding of individual demonstration interviews as research instruments in physics education. Even with in-depth, open-ended questioning, can we be sure of student understanding? When epistemological stance prevents our learning more from a student, we should be aware that we are learning less about the student's difficulties with the physics than about their guiding epistemology.